\documentclass[conference, letterpaper]{IEEEtran}
\IEEEoverridecommandlockouts

\usepackage[top=0.75in, bottom=1in, left=0.625in, right=0.625in]{geometry}

\usepackage{cite}
\usepackage{amsmath,amssymb,amsfonts}
\usepackage{algorithmic}
\usepackage{textcomp}
\usepackage{xcolor}
\def\BibTeX{{\rm B\kern-.05em{\sc i\kern-.025em b}\kern-.08em
    T\kern-.1667em\lower.7ex\hbox{E}\kern-.125emX}}
\usepackage{soul}
\usepackage{svg}
\usepackage{graphicx}
\usepackage{enumitem}
\usepackage{comment}
\usepackage{booktabs} 
\usepackage[caption=false]{subfig}
\usepackage{url}
\usepackage{tikz}
\usepackage{adjustbox}
\usepackage{rotating}
\usepackage{multirow}
\usepackage{lineno}
\usepackage[misc]{ifsym}

\usepackage{dirtytalk}
\usepackage[normalem]{ulem}
\usepackage{blindtext}
\usepackage{tabularx}
\usepackage{tabulary}
\usepackage{wrapfig}
\usepackage{tikz}
\usetikzlibrary{shapes.geometric, arrows}
\usepackage{float}
\usepackage[ruled,vlined]{algorithm2e}
\usepackage{tikz}
\newcommand*\circled[1]{\tikz[baseline=(char.base)]{
            \node[shape=circle,draw,inner sep=1pt] (char) {#1};}}

\usepackage[symbol]{footmisc}
\AtBeginDocument{%
  \providecommand\BibTeX{{%
    \normalfont B\kern-0.5em{\scshape i\kern-0.25em b}\kern-0.8em\TeX}}}

\begin{document}

\title{\Huge{Exploring Jamming and Hijacking Attacks for Micro Aerial Drones}}

\author{
  \IEEEauthorblockN{Yassine Mekdad\textsuperscript{*}, Abbas Acar\textsuperscript{*}, Ahmet Aris\textsuperscript{*}, Abdeslam El Fergougui\textsuperscript{†}, Mauro Conti\textsuperscript{‡},\\ Riccardo Lazzeretti\textsuperscript{§}, and Selcuk Uluagac\textsuperscript{*}}
  \IEEEauthorblockA{\textsuperscript{*}Cyber-Physical Systems Security Lab, Florida International University, Miami, USA\\
  \textsuperscript{†}Moulay Ismail University of Meknes, Meknes, Morocco\\
  \textsuperscript{‡}University of Padua, Padua, Italy\\
  \textsuperscript{§}Sapienza University of Rome, Rome, Italy\\
      Email: \textsuperscript{*}{\{ymekdad, aacar001, aaris, suluagac\}}@fiu.edu, \textsuperscript{†}a.elfergougui@umi.ac.ma,\\ \textsuperscript{‡}mauro.conti@unipd.it, \textsuperscript{§}lazzeretti@diag.uniroma1.it}
}
\maketitle

\begin{abstract}
Recent advancements in drone technology have shown that commercial off-the-shelf Micro Aerial Drones are more effective than large-sized drones for performing flight missions in narrow environments, such as swarming, indoor navigation, and inspection of hazardous locations. Due to their deployments in many civilian and military applications, safe and reliable communication of these drones throughout the mission is critical. 
The Crazyflie ecosystem is one of the most popular Micro Aerial Drones and has the potential to be deployed worldwide. In this paper, we empirically investigate two interference attacks against the Crazy Real Time Protocol (CRTP) implemented within the Crazyflie drones. In particular, we explore the feasibility of experimenting two attack vectors that can disrupt an ongoing flight mission: the jamming attack, and the hijacking attack. Our experimental results demonstrate the effectiveness of such attacks in both autonomous and non-autonomous flight modes on a Crazyflie 2.1 drone. Finally, we suggest potential shielding strategies that guarantee a safe and secure flight mission. To the best of our knowledge, this is the first work investigating jamming and hijacking attacks against Micro Aerial Drones, both in autonomous and non-autonomous modes.

\end{abstract}

\begin{IEEEkeywords}
Unmanned Aerial Vehicles, UAVs, drones, jamming attack, hijacking attack, cybersecurity
\end{IEEEkeywords}

\section{Introduction}
\label{Introduction}

Defense Advanced Research Projects Agency (DARPA) originally introduced the concept of Micro Aerial Vehicle (MAV) for defense applications, with requirements including a maximum wingspan length of 15 cm and a weight of up to 100 g \cite{birth2009}. These micro-sized devices can be carried easily, and they are best suited for specific applications (e.g., indoor navigation, warehouse operations, drone swarming). Recently, many companies, government laboratories, and researchers from academia have been active in the design and development of MAVs~\cite{boroujerdian2022role,kumar2017opportunities}. The Crazyflie quadcopter is one of the most popular Micro Aerial Drones and is ideal for applications in different fields, and swarming on a larger scale~\cite{giernacki2017crazyflie}. However, the threats posed by the nature of wireless communications at the physical layer trigger a series of interference attacks, which are the most common types of attacks targeting MAVs~\cite{Mekdad2021AUAVs}. Indeed, the limited resources of such devices make it very challenging to integrate software or hardware-based security solutions. In this paper, we empirically explore two interference attacks against the CRTP protocol that manages radio communication over the physical layer.
Specifically, we investigate whether the jamming and hijacking attacks hold for the Crazyflie drones in autonomous and non-autonomous flight modes. First, we perform constant jamming of the signal transmission by continuously broadcasting high-power interference signals~\cite{pirayesh2022jamming}. Second, we perform a hijacking attack, in which we aim to 
control the Crazyflie drone. Then, we validate the effectiveness of each attack to gain insight into real-world applications. 
Finally, we suggest potential defense mechanisms and show the challenges that must be considered while integrating such solutions in practical settings.

\noindent \textbf{\textit{Difference from existing work:}} Different from prior works, we perform practical and cost-efficient interference attacks against Micro Aerial Drones, focusing on the CRTP communication protocol. These drones have several unique features that distinguish them from other types of off-the-shelf drones in terms of security and performance. With respect to the existing literature that targets only medium and small drones, this is the first work focusing on interference attacks on Micro Aerial Drones. From a security standpoint, Micro Aerial Drones have limited wireless capabilities as one can suspect their vulnerability at the physical layer.
Thus, the need to provide comprehensive security investigations for such types of drones. On the other hand, our study considers both autonomous and non-autonomous flight modes to analyze our wireless attacks, whereas most of the prior studies have focused only on the non-autonomous flight mode.

\noindent \textbf{\textit{Contributions:}} The main contributions of our work are three-fold:
\begin{itemize} 
    \item We experiment the feasibility of two wireless attacks against the CRTP communication protocol: 
    (i) the jamming attack, and (ii) the hijacking attack.
    \item In a controlled and safe environment, we showcase the real-world consequences of jamming and hijacking attacks and how they can be performed in practical settings. Moreover, we plan to publicly release our research artifacts for the transparency and reproducibility of our results.
    \item We suggest potential shielding strategies to mitigate the security risks posed by the attacks. Then, we  
    discuss the various challenges regarding implementing such solutions.
\end{itemize}
\section{Background}
\label{Background}

In this section, we present the background information on the Crazyflie platform and its communication protocol.

\subsection{Crazyflie UAV Development Platform}
 The Crazyflie platform consists of the \textit{Crazyflie drone}, the \textit{Ground Control Station} (GCS), and the \textit{Crazyradio module}. 
The Crazyflie drone is connected to the GCS through the Crazyradio module. This module not only facilitates the exchange of control and telemetry data between the drone and the GCS but also has broadcasting capabilities for multiple Crazyflie drones communicating on the same radio channel. The Crazyflie ecosystem offers a wide range of features through its client software for the GCS, including real-time logging data, setting flight parameters, and command-based flight control. This enables the Crazyflie drone to be easily integrated into various applications. The availability of the Crazyradio module, the integrated IMU and pressure sensor, and the feature-rich client software for the GCS make the Crazyflie drone an attractive option for deployment in various settings.

\subsection{Crazy Real Time Protocol (CRTP)} The exchanged control and telemetry data between the GCS, and the Crazyflie drones are encoded into CRTP packets and sent via the radio through the ESB protocol. The CRTP protocol enables packet ordering for real-time control of the Crazyflie drone. We note that the payload of the CRTP packets provides the data buffer of the message (up to 31 bytes). The header of the CRTP packets is divided into three segments: \textit{(i) the port}, \textit{(ii) the link}, and \textit{(iii) the channel number}. The port segment ranges between 0 and 15 (4 bits) and identifies the functionality of the message (e.g., port number 3 defines the commander and port number 5 for data logging). The link segment is a reserved field for future use and ranges between 0 and 3 (2 bits), while the channel number ranges between 0 and 3 (2 bits), and is used to determine the sub-functionality of the message that is defined in the application layer. As part of the CRTP protocol, each drone has a Unique Radio Identifier (URI) that is handled by the Crazyradio module. It includes four segments described as follows:

\begin{itemize}
    \item Medium: Defines the nature of communication between the Crazyflie drone and the GCS (e.g., radio, serial).
    \item Radio Channel: Identifies the radio channel between the GCS and the Crazyflie drone that is operating (i.e., the radio channel operates from 0 to 125). 
    \item Communication speed: Refers to the bandwidth during the Air-2-Ground communication and operates under three communication data rates: 250kbit/s, 1Mbit/s, and 2Mbit/s.
    \item Address: Determines the radio address of the Crazyflie drone communicating with the GCS. We mention that the default radio address of the Crazyflie drone provided by the manufacturer is \texttt{E7:E7:E7:E7:E7} and can be changed by the operator.
\end{itemize}
\section{Threat Model}
\label{ThreatModel}
\begin{figure}[!h]
\begin{center}
\includegraphics[width=\columnwidth]{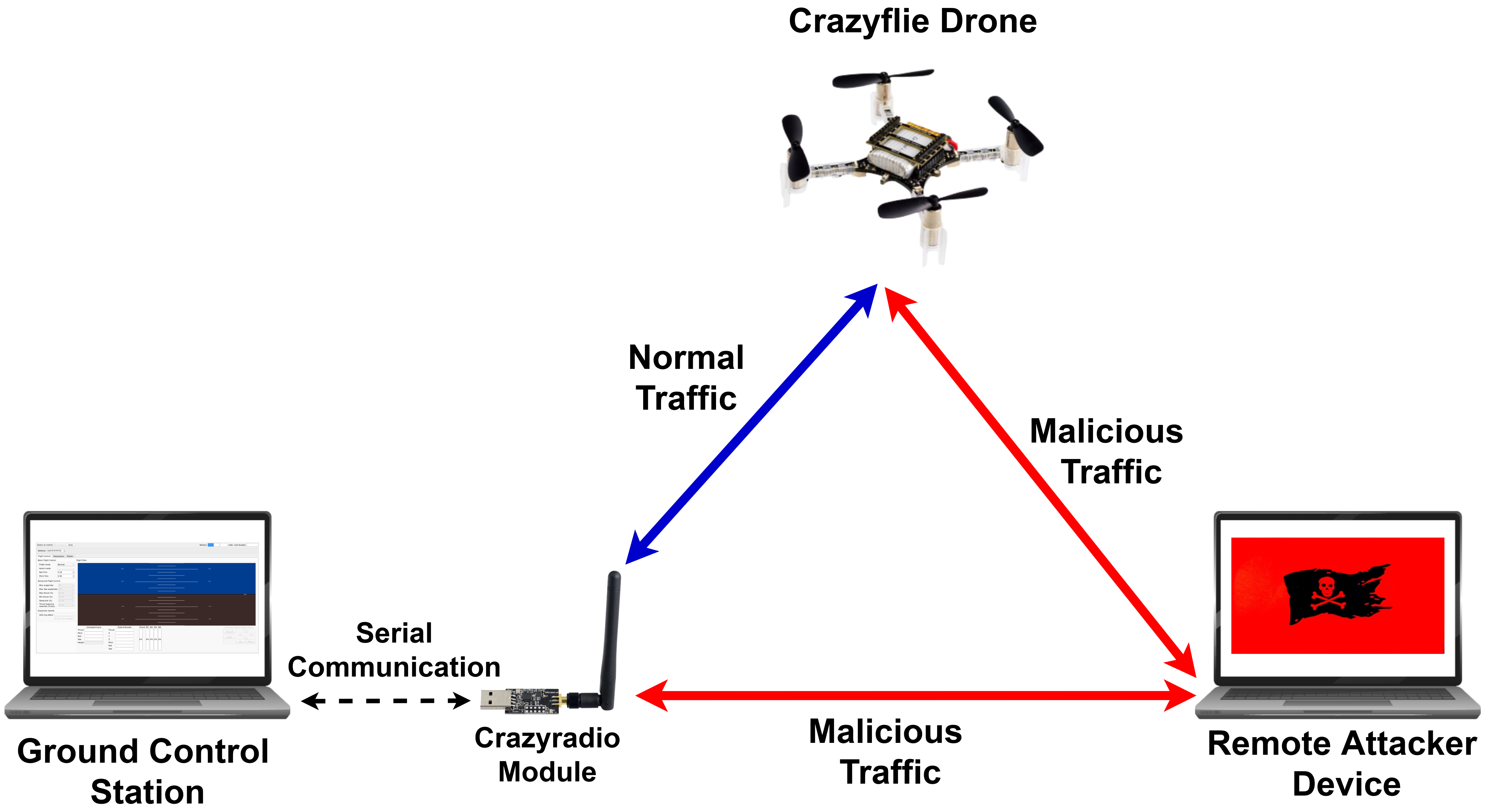}
\caption{The wireless threats considered in a Crazyflie ecosystem.}
\label{Threat}
\end{center}
\end{figure}

In our study, we consider an attacker within the range of the radio space of the drones (i.e., 1 km range line-of-sight with the Crazyflie drone) without physical access. The attacker's goal is to compromise the flight mission. As illustrated in Figure~\ref{Threat}, the adversary can remotely transmit malicious traffic from his device to the Crazyflie drone and the Crazyradio module. Here, we consider an active attacker who can prevent the Air-2-Ground communication link, thus threatening the availability of the flight mission by taking over the drone mid-flight through a jamming or a hijacking attack. Regarding the attacker capabilities, we assume that the adversary can purchase similar or identical devices (e.g., Crazyflie drone, Crazyradio module) to develop and test the attacks. 

\section{Attack Implementation and Evaluation}
\label{Attack}
In this section, we present our experimental setup. Then, we provide the implementation of our attacks. Afterward, we discuss the practicality and cost analysis of our attacks. 

\subsection{Experimental Setup} \label{subsec:setup}
In our experiments, we consider a centralized network architecture~\cite{Mekdad2021AUAVs}, where we used one Crazyflie 2.1 drone flashed with the latest firmware version (v2022.09), which communicates with the GCS through the Crazyradio Power Amplifier module (i.e., USB dongle~\cite{carzyradio}). The GCS is a laptop computer with an Intel Core i7-9750H CPU clocked at 2.60 GHz and 16 GB of memory. We note that the Crazyradio module is also flashed with the latest firmware version (v0.53). First, we setup the radio address of the Crazyflie drone on \texttt{01:E7:E7:E7:E7}. Then, we setup the Crazyradio module on channel 81 with a 2 Mbits/s radio bandwidth.

\subsection{Attack Implementation} \label{subattack}
To demonstrate the feasibility and effectiveness of our attacks
against the Crazyflie ecosystem, we performed two physical layer attacks, 
where it is sufficient for the adversary to compromise the 
availability of the ongoing flight mission, and then ultimately take over control of the drone. We performed all our 
attacks against the Crazyflie drone by considering two different flight mode scenarios: (i) autonomous flight mode and (ii) non-autonomous flight mode. In the autonomous flight mode, the Crazyflie drone operates autonomously via a client Python-based API~\cite{python}. In the non-autonomous flight mode, the user manually operates the Crazyflie drone via a real-time command-based flight control. It is worth mentioning that regardless of the considered attacks, the scanning phase is a prerequisite to gathering the radio address and the operating channel of the flying drone.\\

\noindent \textbf{Scanning:} To target a specific drone, we first start by scanning the network to identify the active drones and their communication channels. 
Therefore, in the scanning phase, we look for connected Crazyflie drones promiscuously by sweeping all possible channels under different radio bandwidths. Then, we display the active radio addresses of the Crazyflie drones with their corresponding radio channels and the received packets. 
According to Shannon's information theory, we mathematically quantify the amount of data scanned $I_s$. In fact, given a continuous Air-2-Ground communication system, the transmitted signal $x(t)$ from the GCS to the drone can be modeled as a continuous signal with a probability density function $p(x)$, where its entropy density $H$ can be defined as follows:
\begin{equation}
H(X) = -\int_{-\infty}^{\infty} p(x) \log_2 p(x) dx
\end{equation}

Similarly, the received signal $y(t)$ by the drone from the GCS can be described as a continuous signal with a probability density function $p(y)$, with its entropy density $H(Y)$.
can be defined as:
\begin{equation}
H(Y) = -\int_{-\infty}^{\infty} p(y) \log_2 p(y) dy
\end{equation}
To quantify the amount of information shared between the drone and the GCS, we define $I(X:Y)$ as the mutual information between the transmitted signal and the received signal, which can be expressed as:

\begin{equation}
I(X:Y) = \int \int p(x,y) \log_2 \left(\frac{p(x,y)}{p(x)p(y)}\right) dx dy
\end{equation}
where $p(x,y)$ is the joint probability distribution of the transmitted and received signals $x(t)$ and $y(t)$, respectively. Here, the scanning device aims to gather information about the transmitted signal $x(t)$ by observing the received signal $y(t)$. To determine the amount of data scanned, we subtract the entropy density of the transmitted signal from the mutual information as follows:
\begin{equation}
I_s = I(X:Y) - H(X)
\end{equation}

\begin{figure}[!h]
\begin{center}
\includegraphics[width=\columnwidth]{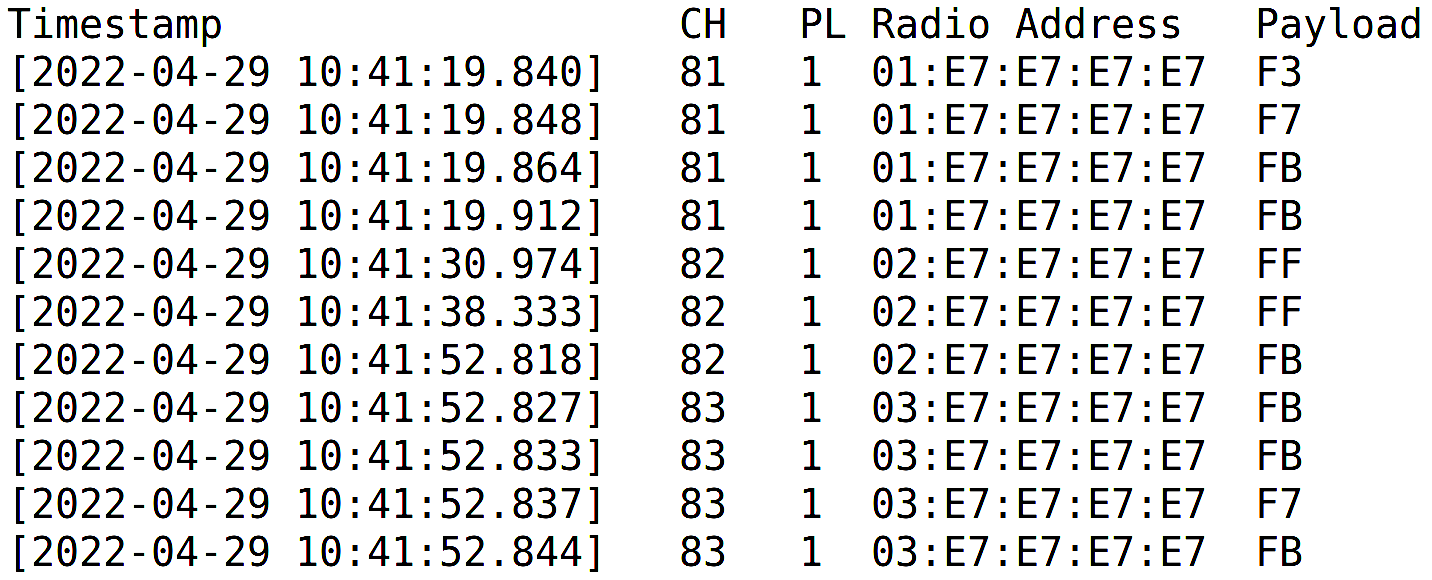}
\caption{An example of scanning three Crazyflie drones using the Crazyradio-sniffer. CH refers to the radio channel, and PL refers to the payload length.}
\label{Scanning}
\end{center}
\end{figure}

Figure~\ref{Scanning} illustrates the result of the scanning phase by considering three active drones with the radio addresses \texttt{01:E7:E7:E7:E7}, \texttt{02:E7:E7:E7:E7}, and \texttt{03:E7:E7:E7:E7} operating on three different channels: 81, 82, and 83, respectively. For the sake of clarity, we set the target drone on channel 81. We note that network scanning cannot be simply prevented as the device identifier and channel information data is inherently available to the public, similar to the MAC address and channel information in WiFi. It is worth mentioning that regardless of the implemented attacks, we scan the network to gather the radio address and the operating channel of the flying drone. We note that network scanning is a base for all implemented attacks in our study.

\begin{figure}[!h]
\centering
\includegraphics[width=0.9\columnwidth]{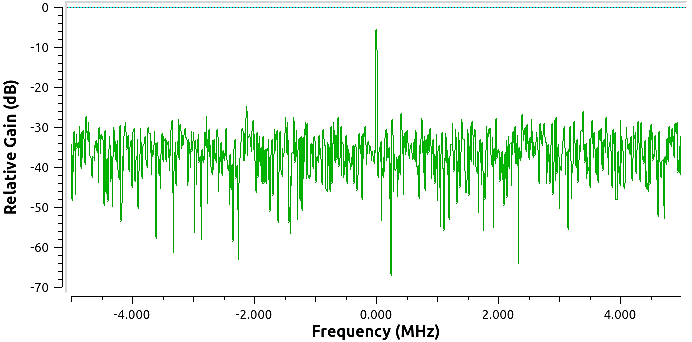}
\caption{ Frequency sink flow graph for a Crazyflie drone operating on a radio frequency range centered in 2481 MHz.}
\label{Frequency}
\end{figure}

\smallskip
\noindent \textbf{Attack 1 - Jamming:} We launched the jamming attack against the running Crazyflie drone by jamming the signal going to and coming from the GCS. This attack consists of blocking the Air-2-Ground communication link and consequently disrupting the availability of the flight mission. Given a wireless communication system transmitting a benign signal $x(t)$ to a receiver over a period of time $t$. The adversary transmits jamming signal $j(t)$, which is added to the benign signal, such that 
\begin{equation}\label{equation}
y(t) = x(t) + j(t)
\end{equation}
where $y(t)$ is the received signal. We mention that the jamming signal is effective if and only if the received signal is corrupted. In other words, the jamming signal can be described as an additive noise that corrupts the benign signal. To evaluate the effectiveness of the corrupted signal, the adversary decreases the Signal-to-Noise Ratio (SNR) (i.e., the metric that measures the quality of the received signal). The SNR is the ratio of signal power to the noise power and can be defined as follows:

\begin{equation}
SNR = 10 \log_{10} \left(\frac{||x(t)||^2}{||j(t)||^2}\right)
\end{equation}
such that $||.||$ refers to the Euclidean norm of the signal. Given the affordability of low-cost Software Defined Radios, we performed the jamming attack using HackRF One, one of the commercially available Software Defined Radios, and its open-source software development toolkit such as GNU radio. In the literature, there exist several types of jamming attacks, such as constant, reactive, or random jamming attacks~\cite{pirayesh2022jamming}. In our case, we conducted a constant jamming attack, where we broadcast high-power interference signals (i.e., Gaussian noise) continuously over the operating radio frequency. We mention that the operating frequency of the Crazyflie drone can be inferred from its corresponding radio channel. For example, the Crazyflie drone operating on radio channel 81 has a radio frequency of 2481 MHz (i.e., the sum of 2.4 GHz and 81 MHz). We illustrate in Figure~\ref{Frequency} the captured frequency sink flow graph of a Crazyflie drone with a radio frequency range centered at 2481 MHz. We set the frequency bandwidth to 10 MHz and the Fast Fourier Transform (FFT) size to 1024. The pick of the relative gain from the center frequency can be clearly shown in Figure~\ref{Frequency}, which means that we can perfectly transmit interference signals to prevent the Crazyflie drone from accessing the radio channel.

\begin{figure}[!h]
\centering
\includegraphics[width=0.9\columnwidth]{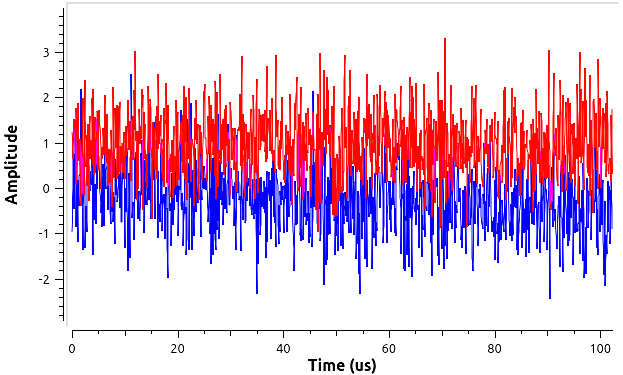}
\caption{Flow graph of the transmitted jamming signal over a radio frequency of 2481 MHz. The amplitude is a unitless measure that presents the standard deviation of the signal from the mean. The blue color and the red color refer to the real and imaginary parts of the transmitted complex signal, respectively.}
\label{Time}
\end{figure}

After gathering the radio frequency of the Crazyflie drone, we continuously broadcast a Gaussian noise signal over the radio channel to jam the Air-2-Ground communication. In this case, the transmitted signal is in the form of a complex source that is divided into real and imaginary parts. We consider a constant jamming signal that is characterized by a high sampling rate of up to 10 million samples per second (i.e., 10MHz) to guarantee high accuracy in capturing and processing the signal. We amplify the jamming signal by setting the Radio Frequency (RF) gain to 14 dB, and we believe that the strength of our RF gain is fair enough to make the legitimate signal from the malicious one indistinguishable. To adequately filter out the generated jamming signal, we set a high cutoff frequency of 4MHz, with a transition width of 1MHz. In the Intermediate Frequency (IF) domain, we consider the IF gain of 47dB and the baseband gain of 0 dB. The working of the transmitted jamming signal with an amplitude of 1 is shown in Figure~\ref{Time}. After performing the jamming attack against the Crazyflie drone in both autonomous and non-autonomous flight modes, we observe that the GCS cannot establish Air-2-Ground communication. Moreover, the Crazyflie drone crashes down in the autonomous flight mode while it remains stable over the air in the non-autonomous flight mode.

\begin{figure}[!h]
\centering
\includegraphics[width=0.99\columnwidth]{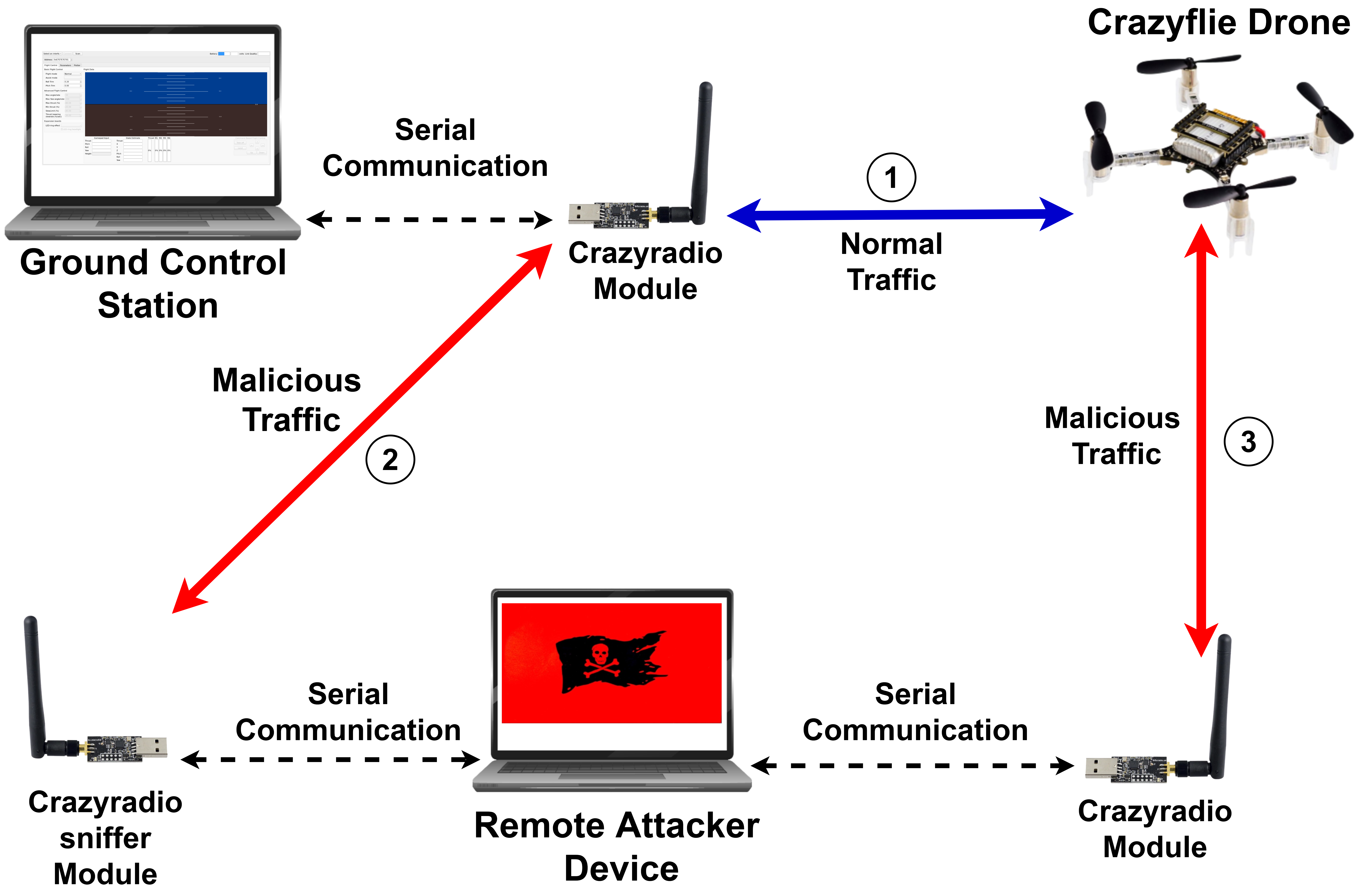}
\caption{Scenario of hijacking attack against the Crazyflie drone.}
\label{Hijacking}
\end{figure}

\noindent \textbf{Attack 2 - Hijacking:} To perform the hijacking attack against the Crazyflie drone during the flight mission, we break the ongoing Air-2-Ground communication link. Then, we take complete control of the drone. The hijacking attack can be described similarly to the jamming attack, except the goal of the adversary is to transmit a hijacking (i.e., unauthorized) signal $h(t)$ that aims to mimic the transmitted signal. Thereby, fooling the receiver (e.g., drone) into executing unintended instructions. Therefore, the hijacking attack over a period of time $t$ can be defined as follows:

\begin{equation}
y(t) = x(t) + h(t)
\end{equation}
such that $y(t)$, $x(t)$, and $h(t)$ are the received signal, the transmitted signal, and the hijacking signal, respectively. By using the previously mentioned Crazyradio-sniffer module and another Crazyradio module, we can perfectly hijack the Crazyflie drone. The remote attacker device has a similar configuration as the GCS. Here, we considered additional off-the-shelf equipment connected to the remote attacker device for each wireless attack (e.g., Crazyradio Power Amplifier module, HackRF One~\cite{hackrf}). We illustrate the hijacking attack in Figure~\ref{Hijacking}. First, we run the Crazyflie drone and establish the Air-2-Ground communication link \circled{\textbf{1}}. We can observe the normal behavior of the flight mission from the GCS side in both autonomous and non-autonomous flight modes. Afterward, we perform a single-tone attack through the Crazyradio-sniffer module~\cite{sathaye2019wireless}. In particular, we transmit a continuous single-tone signal on the frequency of the targeted radio channel, where the signal strength of the Crazyradio-sniffer module is higher than the legitimate Crazyradio module \circled{\textbf{2}}. As a result, potential interference is likely to happen, and the GCS can no longer send or receive the packets from the drone. It is worth mentioning that transmitting a continuous single-tone signal at the same frequency as the targeted radio channel (also known as Continuous Wave (CW) jamming) will likely produce a jamming signal (i.e., potential signal interference with the legitimate Air-2-Ground communication link). Moreover, the drone is disconnected from the GCS. By taking advantage of this behavior, we eventually hijack the Crazyflie drone by establishing our malicious Air-2-Ground communication link through a second Crazyradio module \circled{\textbf{3}}. When hijacking the Crazyflie drone in both autonomous and non-autonomous flight modes, we notice that the GCS can no longer control the drone. On the other hand, the Crazyflie drone crashes down in autonomous flight mode. In this case, we can physically access the targeted drone in our specified area (e.g., beyond line-of-sight). In the non-autonomous flight mode, the Crazyflie drone remains stable over the air, and we can completely take control of the flying drone, thus enabling a smooth hijacking attack. Another type of hijacking happens when the Crazyflie drone is active and waiting for a connection from the user. In this case, we cannot distinguish between a malicious and a legitimate user since both of them can establish the GCS-2-UAV communication without preliminary authentication.

Under the \textit{autonomous flight mode}, the Crazyflie drone crashes in the jamming and hijacking attack. This behavior can be attributed to the absence of a safe-mode feature in the Crazyflie drone, which could have prevented potential accidents for the drone and its surroundings. In this scenario, when the adversary disrupts the programmed instructions through jamming or hijacking attacks, the Air-to-Ground communication link is lost, resulting in the crash of the Crazyflie drone. In the \textit{non-autonomous flight mode}, the jamming and hijacking attacks against the Crazyflie drone exhibit different behavior than the autonomous flight mode. Since the Crazyflie drone is manually operated by the user in the Line-of-Sight range, the control commands are periodically transmitted from the GCS to the Crazyflie drone. In this case, whenever the Crazyflie drone is under jamming or hijacking attack, the user cannot transmit control commands to the flying drone. Consequently, the drone holds the last command and remains in a suspended state until receiving a new control command from the user.

\subsection{Attack Practicality and Cost Analysis} \label{practicality}

\noindent\textbf{Attack Demonstration.} We made a video\footnote{\url{https://www.youtube.com/watch?v=kgG9b3UmoW0}} in a safe and controlled environment, to show the effectiveness and consequences of our attacks against the Crazyflie drone in the real world. In this demonstration, we considered the experimental setup shown in Section \ref{subsec:setup}. First, we 
demonstrate the possibility of preventing the Air-2-Ground communication link by constantly jamming the signal transmission. Then, we prove that the Crazyflie drones are subject to hijacking attacks by crashing them down on a specific location or maliciously controlling them. 

\noindent\textbf{Cost Analysis.} We list the required hardware and approximate total cost for each physical layer attack in Table~\ref{Hardware}. Overall, all of these attacks can be performed by commercial-of-the-shelf low-cost devices.
\begin{table}[!h]
\centering
\caption{Required hardware and approximate cost for the wireless attacks against the Crazyflie drone}
\label{Hardware}
\def\arraystretch{1.2}
\resizebox{1.08\columnwidth}{!}{
\begin{tabular}{llclll}
\cline{1-4}
\multicolumn{1}{|l|}{\textbf{Attack Name}}            & \multicolumn{1}{l|}{\textbf{Required Hardware}}                                                                                                                   & \multicolumn{2}{c|}{\textbf{Total Cost}}                                                                           &  &  \\ \cline{1-4}
\multicolumn{1}{|l|}{\multirow{3}{*}{Jamming Attack}} & \multicolumn{1}{l|}{\multirow{3}{*}{\begin{tabular}[c]{@{}l@{}}-Crazyradio PA 2.4 GHz \\  -HackRF One (SDR) \\ -ANT500 Telescope Antenna\end{tabular}}} & \multicolumn{1}{c|}{\$35.00}  & \multicolumn{1}{l|}{\multirow{3}{*}{\$410.00}} &  &  \\ \cline{3-3}
\multicolumn{1}{|l|}{}                                & \multicolumn{1}{l|}{}                                                                                                                                             & \multicolumn{1}{c|}{\$340.00} & \multicolumn{1}{l|}{}                         &  &  \\ \cline{3-3}
\multicolumn{1}{|l|}{}                                & \multicolumn{1}{l|}{}                                                                                                                                             & \multicolumn{1}{l|}{\$35.00}  & \multicolumn{1}{l|}{}                         &  &  \\ \cline{1-4}
\multicolumn{1}{|l|}{Hijacking Attack}                & \multicolumn{1}{l|}{-Two Crazyradio PA 2.4 GHz}                                                                                                                   & \multicolumn{2}{c|}{\$70.00}                                                  &  &  \\ \cline{1-4}
                                                      &                                                                                \end{tabular}
}
\end{table}
\section{Potential Defense Solutions and Challenges}

\label{Countermeasures}  
\subsection{Defense Solutions}
To effectively address the implemented attacks against the Crazyflie ecosystem, we present below potential mitigation strategies enabling a secure Crazyflie-based flight mission.

\subsubsection{Jamming Attack} To avoid jamming attacks against the Crazyflie ecosystem, Intrusion Detection Systems (IDS) should be implemented onboard or on the GCS side to detect different categories of malicious traffic (e.g., message forgery attacks, routing attacks, signals modification)~\cite{choudhary2018intrusion}. Other defense mechanisms at the physical layer can be mitigated by utilizing anti-jamming strategies (e.g., channel/frequency hopping techniques, Direct Sequence Spread Spectrum (DSSS) techniques, Multiple-Input and Multiple-Output (MIMO) techniques, etc.)~\cite{pirayesh2022jamming}. However, there is no unified solution that can be effective against all classes of jamming attacks.    

\subsubsection{Hijacking Attack} To defend against hijacking attacks, efficient hijacking detection mechanisms need to be implemented either into the Crazyflie drone or the GCS (e.g., statistical analysis of the flight patterns, GPS-based detection methods, estimated position techniques based on onboard Inertial Measurement Unit (IMU))~\cite{feng2018efficient}. These solutions help prevent an adversary from physical theft and taking control of the drone during the flight mission. Another defense strategy to protect the Crazyflie drones from jamming and hijacking attacks consists of enabling and implementing the \textit{safe mode} feature. This feature allows the drone to automatically follow a set of pre-programmed instructions in the case of a loss of communication with the GCS or failure at various levels, such as hardware, software, and communication. These instructions ensure the safety and security of the drone. Examples of actions that could be included in the safe mode include emergency landing, returning home, and sending GPS location. 

\subsection{Challenges}
\label{Challenges}  

The popularity of commercial-off-the-shelf drones in the past few years with their civilian and military applications created a wide range of challenges~\cite{nassi2021sok}. Some of these challenges include the lack of standardization and unified solutions, performance considerations, and usability-centered design. 

\noindent \textbf{Lack of Standardization and Unified Solution.} With the fast-growing deployment of brand-new commercial off-the-shelf drones, it is becoming increasingly difficult to guarantee the confidentiality and availability of the flight mission. These challenges could be explained due to the lack of standardization and unification regarding the architectural design of commercial drones~\cite{valente2017understanding}. On the other hand, there is no unified solution that prevents all three attacks at once. For example, establishing a secure Air-2-Ground communication channel can likely prevent an eavesdropping attack. However, it does not stop jamming or hijacking attacks. Therefore, each attack requires a unique defense to be implemented, which makes it difficult to integrate into real commercial off-the-shelf drones.

\noindent \textbf{Performance Considerations.} It is becoming challenging to implement security solutions without decreasing the performance of the drone system~\cite{mekdad2023comprehensive}. Adopting cryptography-based approaches might require additional computation costs and potentially increase energy consumption~\cite{oz2023rob}. For instance, the deployment of an onboard IDS will be power-consuming. In contrast, its deployment in the GCS will negatively impact the performance of the Air-2-Ground communication link by introducing end-to-end delays. Although hijacking detection solutions can effectively prevent physical theft, the recent release of anti-drone products in the market can be used for malicious purposes (e.g., drone guns, drone spoofers), and consequently neutralizing legitimate drones~\cite{chamola2021comprehensive}.

\noindent \textbf{Usability-centred Design.} Existing manufacturers avoid incorporating security features when designing commercial drones. Although integrating security features can improve the overall security of  
the drones, these features can also negatively impact the usability of the drones in terms of performance (e.g., communication cost, computation cost, energy consumption). In fact, the primary business goal of these manufacturers is to design lightweight and user-friendly drones that are capable of meeting the needs of their target market. However, the trade-off between usability and security when designing and deploying commercial drones is challenging, as it requires a balancing strategy that can provide an acceptable level of security while also maintaining a user-friendly experience~\cite{cayir2023augmenting}.
\section{Conclusion}
\label{Conclusion}
In this paper, we empirically studied and analyzed the physical layer security of Micro Aerial Drones. In particular, we demonstrated the feasibility of performing the jamming, and the hijacking attack in both autonomous and non-autonomous flight modes for Crazyflie drones. Our experimental results show that the root cause of such attacks comes from the inherent design of the Crazyflie drones' physical layer. To address such issues, we suggested a set of potential defense mechanisms for each attack vector, which could be implemented within the CRTP communication protocol. 
Finally, we discussed the challenges that need to be considered during the implementation of the potential defense mechanisms. Future research efforts will be devoted to incorporating authentication mechanisms within the CRTP protocol.

\section{Acknowledgment}
This work was partially supported by the US National Science Foundation (Awards: 2039606, 2219920), Florida International University Graduate School, and Microsoft. The views expressed are those of the authors only, not of the funding entities.

\bibliographystyle{IEEEtran}
\bibliography{References}

\end{document}